\documentclass[10pt,a4paper,tightenlines,twocolumn,prd,showpacs,showkeys,lengthcheck,nobibnotes]{revtex4}
\usepackage{amsfonts}
\usepackage{amsmath}
\usepackage{amssymb}
\usepackage{epsfig}
\usepackage{graphicx}
\usepackage{fancyhdr}

\begin{document}

\preprint{}
\title[ ]{\textbf{Stage analysis of delayed-choice and quantum eraser
experiments}}
\author{George Jaroszkiewicz}
\affiliation{University of Nottingham, University Park, Nottingham, NG7 2RD, UK}

\begin{abstract}
Delayed choice and quantum eraser experiments have attracted much interest
recently, both theoretically and experimentally. In particular, they have
prompted suggestions that quantum mechanics involves acausal effects. Using
a recently developed approach which takes apparatus into account, we present
a detailed analysis of various double-slit experiments to show that this is
never the case. Instead, quantum experiments can be described in terms of a
novel concept of time called stages. These can cut across the conventional
linear time parameter as experienced in the laboratory and appear to violate
causality.
\end{abstract}

\maketitle

\section{Introduction}

Delayed-choice \cite{WHEELER-1978,JACQUES+AL-2006}, quantum eraser \cite%
{WALBORN-2002} and delayed-choice quantum eraser \cite{KIM+AL-2000}
experiments\ have led to suggestions that interference patterns formed by
particles impacting on a screen may be influenced in some way by decisions
made long after those particles had landed on that screen. Our objective in
this paper is to show by a detailed analysis of various experiments that
quantum principles do not support those suggestions.

In our analysis, we shall make out a case for the adoption of a perception
of time in quantum observation different to that used in classical
mechanics. To understand what we mean, it is important to keep a clear
distinction between the concepts of \emph{systems under observation} (SUOs),
such as photons, and \emph{apparatus}. Our formalism will reflect this
difference consistently. Standard unitary Schr\"{o}dinger evolution can be
maintained for states of SUOs in between preparation and outcome, but
apparatus appears to follows different rules.

Recent quantum experiments \cite%
{WALBORN-2002,JACQUES+AL-2006,KIM+AL-2000,KIM-2003} are consistent with and
support the view that the passage of $``$detector time$"$ is synonymous with
quantum information acquisition occurring in a sequence of \emph{stages}
\cite{J2005A}. Stages have rules which are not precisely those of classical
information acquisition, and it appears to be this which accounts for much
of the well-known difficulty we have in explaining on a classical level
various quantum mechanical experiments involving quantum interference.

These rules conform with known physics. For example, quantum information
acquisition never violates the light-cone constraints of relativity:
classical information cannot be acquired between spacelike intervals.
Quantum correlations which appear to violate the principle of Einstein
locality actually always require observations to be completed before those
correlations can be defined, and this completion always takes place in a
classically consistent matter.

Another rule is that quantum information in the form of SUO states \emph{can}
be shielded against the effects of decoherence and preserved in a state of
stasis for arbitrarily long periods of laboratory time. This is most evident
in the Heisenberg picture and is confirmed by the observation of light from
distant stars and galaxies. It is also one way to understand particle decay
experiments \cite{J2007G}.

Yet another rule is that the observation of different components of
entangled states is best discussed in terms of stages rather than linear
laboratory time. This rule is responsible for the apparent acausality in the
delayed-choice experiments we are interested in: observations involving
separate detectors can be taken in apparently random order relative to
laboratory time without affecting correlations. This was confirmed in the
case of the double-slit quantum eraser experiment by Walborn et al. \cite%
{WALBORN-2002}, who specifically looked at this issue.

Our approach uses a formalism that we have developed for the analysis of
time-dependent quantum apparatus networks \cite{J2007C}. We have recently
applied it to the Franson-Bell experiment \cite%
{FRANSON-1989,KWIAT-1993,J2008E}, an experiment that appears to involve
acausal quantum interference.

In contrast to standard approaches which tend to focus on the quantum
mechanics of systems under observation (SUOs) such as photons, our approach
focuses on the detecting apparatus as well. This permits a stage-by-stage
analysis of the processes involved in typical quantum optics experiments,
starting from state preparation, through the various modules making up the
apparatus and ending up with the final state detectors. The formalism is
particularly good at giving coincidence rates, which are crucial to many
recent quantum optics experiments such as the delayed-choice quantum eraser.

Our notation serves two purposes. First, it provides an efficient method for
dealing with quantum networks of great complexity and can be readily encoded
into computer algebra packages. Second, it distinguishes between the quantum
states of apparatus detectors and those conventionally associated with SUOs
such as photons. The formalism is completely consistent with all standard
quantum principles.

We make a number of standard assumptions throughout our analysis. First,
complete efficiency is assumed, but of course, no real experiment is like
that. However, many experiments show precisely those important quantum
features such as interference bands which idealized discussions such as ours
predict. Because of that, there is no need to introduce environmental
factors such as decoherence into the discussion. There seems to be no need
either to use full-scale quantum field theory in order to draw out the
important features of the processes we discuss.

In our analysis we shall for economy frequently refer to \emph{photons} as
if they existed in some physical sense as particles. A better interpretation
consistent with our modelling would be in terms of detector signals. Photon
spin is then most naturally interpreted in terms of the specific physical
properties of photon detectors.

\section{The double-slit experiment}

In this section we discuss the double-slit (DS) experiment. This experiment
features prominently as a component of the delayed-choice quantum eraser
experiment and Wheeler's delayed-choice experiment discussed subsequently.

Figure $1$ shows a schematic diagram representing the main features of the
DS experiment. Our notation is as follows. In such a diagram, $A_{n}^{i}$
represents the labstate $\mathbb{A}_{i,n}^{+}|0,n)$ of the $i^{th}$
elementary signal detector located at the indicated place in the apparatus
network at stage $\Omega _{n}$. Here $\mathbb{A}_{i,n}^{+}$ is the
associated signal creation operator and $|0,n)$ is the void state of the
apparatus \cite{J2007C} at stage $\Omega _{n}$.
\begin{figure}[t]
\centerline{\psfig{file=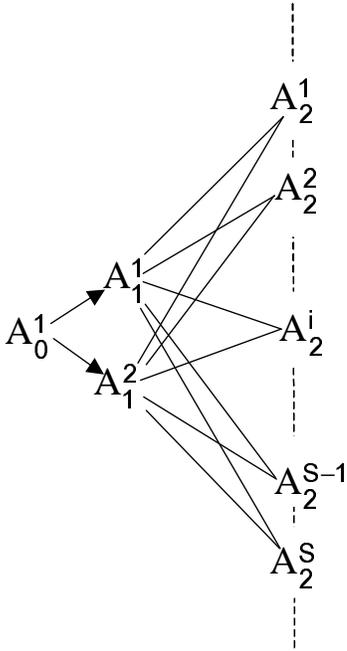,width=2in}}
\caption{The double-slit experiment.}
\end{figure}
A stage can often be identified with a particular moment or period of
laboratory time (i.e., something approximating an instant of simultaneity in
the laboratory rest frame), but this need not be the case at all. In
particular, a given stage during a delayed-choice experiment may involve an
enormous interval of laboratory time, far beyond any notion of simultaneity.
What is crucial in the definition of a stage is that all the detectors
associated with a given stage are \emph{effectively and mutually spacelike}
relative to each other. In other words, no information in any form is
transmitted between them within that stage. Successive stages are defined in
terms of either actual or potential transmission of information between them
in either classical and quantum forms. Exactly what this means will be made
clearer during our discussion of delayed-choice experiments.

In the DS experiment, the source of the photon beam impinging on the
double-slit is denoted by $A_{0}^{1}$ in Figure $1$, the subscript denoting
that it occurs at stage $\Omega _{0}$ and the superscript denoting that
there is a single photon source. The initial labstate is taken to be of the
form
\begin{equation}
\Psi _{0}=\Psi _{0}^{1}s_{0}^{1}A_{0}^{1},  \label{ERASER-01}
\end{equation}%
where $s_{0}^{1}$ is the normalized spin state of the photon concerned and $%
\Psi _{0}^{1}$ is a complex valued normalization factor related to the
initial beam characteristics. In our notation, $\bar{\Psi}_{0}\Psi
_{0}=|\Psi _{0}^{1}|^{2}$.

The initial labstate is not normalized to unity because the formalism
actually determines \emph{relative probability rates} for photon signal
production during relevant photon coherence times and related times
connected with the passage of wave-trains through the apparatus. Related to
this is the requirement for specific contextual information about the
apparatus to be taken into account. For instance, given a very large
detecting screen, some of its detectors would signal photon detection much
earlier on in a given run than other detectors further way from the slits.
Wheeler's delayed-choice experiment is an example where such contextual
information is crucial to the discussion \cite{WHEELER-1978}.

In the basic DS experiment, there are no issues with photon spin, so
actually the $s_{0}^{1}$ term in (\ref{ERASER-01}) is redundant here.
However, photon pair spin is a factor in the double-slit quantum eraser
experiment discussed later on, so we include a photon spin term here to show
how we deal with it in our formalism.

The next step is to compute the effective evolution operator $U_{1,0}$ which
takes the initial labstate from $\Psi _{0}$ to $\Psi _{1}$ in the transition
from stage $\Omega _{0}$ to $\Omega _{1}$. Referring to Figure $1$, we write%
\begin{equation}
U_{1,0}s_{0}^{1}A_{0}^{1}=\alpha ^{1}s_{1}^{1}A_{1}^{1}+\alpha
^{2}s_{1}^{1}A_{1}^{2},
\end{equation}%
where $\alpha ^{1}$ and $\alpha ^{2}$ are complex coefficients satisfying
the rule $|\alpha ^{1}|^{2}+|\alpha ^{2}|^{2}=1$. We do not require at this
point to have a symmetrical double-slit device, so $\alpha ^{1}$ and $\alpha
^{2}$ are not assumed equal in magnitude. Moreover, each slit could induce a
separate phase change in that part of the beam passing through it, so this
is left open as well.

The transition from stage $\Omega _{0}$ to stage $\Omega _{1}$ involves a
change in Hilbert space dimension from $4$ to $8$, according to the counting
rules of our formalism. However, the effective dimension involved with stage
$\Omega _{0}$ is just $1$ and we find%
\begin{equation}
U_{1,0}\backsimeq s_{1}^{1}\bar{s}_{0}^{1}\sum_{a=1}^{2}\alpha ^{a}A_{1}^{a}%
\bar{A}_{0}^{1},
\end{equation}%
where $\bar{s}_{0}^{1}$ and $\bar{A}_{0}^{1}$ are the respective duals of $%
s_{0}^{1}$ and $A_{0}^{1}$ and $\backsimeq $ denotes $``$effective$"$.

The next step is to calculate $U_{2,1}$, the effective evolution operator
from $\Omega _{1}$ to $\Omega _{2}$, at which point we have completed our
dynamical account of the DS experiment. Referring again to Figure $1$, we
write%
\begin{equation}
U_{2,0}s_{1}^{1}A_{1}^{a}=\sum_{i=1}^{S}V^{i,a}s_{2}^{1}A_{2}^{i}\text{, }\
\ \ a=1,2,  \label{ERASER-04}
\end{equation}%
where we have modeled the detecting screen as consisting of a large number $S
$ of photon detectors. This accords with actual experiments, which never
actually involve a continuum of detectors. If necessary, we are free to take
$S$ as large as required in order to model an observed probability pattern
to the accuracy required. Note that the detectors in this collection need
not be assumed to be coplanar: they could be distributed throughout a
three-dimensional region of physical space. Our discussion is perfectly
general.

What keeps a track of total probabilities are the semi-unitarity relations
between the complex coefficients $\{V^{i,a}\}$, which represent the
transition amplitudes from emitters based in $\Omega _{1}$ to detectors in $%
\Omega _{2}$. These relations take the form%
\begin{equation}
\sum_{i=1}^{S}\bar{V}^{i,a}V^{i,b}=\delta _{ab},\ \ \ \ \ a,b=1,2\text{,}
\label{ERASER-02}
\end{equation}%
where $\bar{V}^{i,a}$ denotes the complex conjugate of $V^{i,a}$. Again,
this is perfectly general. There is no need for our purposes to specify any
particular form for the $V^{i,a}$ coefficients, provided the basic
semi-unitarity conditions (\ref{ERASER-02}) are satisfied. It is traditional
in standard discussions of the double-slit experiment to solve the Schr\"{o}%
dinger equation for a monochromatic beam impinging on a screen from two
point sources, but this is not necessary in a discussion such as ours: a
more general formulation reveals the important features of the experiment
more clearly and precisely.

From (\ref{ERASER-04}) we find the effective transition operator $U_{2,1}$
from $\Omega _{1}$ to $\Omega _{2}$ to be given by%
\begin{equation}
U_{2,1}\backsimeq s_{2}^{1}\bar{s}_{1}^{1}\sum_{i=1}^{S}%
\sum_{a=1}^{2}A_{2}^{i}V^{i,a}\bar{A}_{1}^{a}.
\end{equation}%
The total effective evolution operator $U_{2,0}$ is given by the product $%
U_{2,1}U_{1,0}$ and is found to be%
\begin{equation}
U_{2,0}=s_{2}^{1}\bar{s}_{0}^{1}\sum_{i=1}^{S}\sum_{a=1}^{2}\alpha
^{a}A_{2}^{i}V^{i,a}\bar{A}_{0}^{1}.
\end{equation}

The next step is to calculate the generalized Kraus operators $M_{2,0}^{i}$,
defined by%
\begin{equation}
M_{2,0}^{i}\equiv \bar{A}_{2}^{i}U_{2,0}=s_{2}^{1}\bar{s}_{0}^{1}%
\sum_{a=1}^{2}\alpha ^{a}V^{i,a}\bar{A}_{0}^{1},\ \ \ \ \ i=1,2,\ldots ,S.
\label{ERASER-03}
\end{equation}%
In principle, there are $2^{S}$ such operators, counting multiple
coincidence cases, but in the case of the DS experiment, we know we are
dealing with single photon interference. This means that only the $S$
one-signal Kraus operators are non-zero, as given by (\ref{ERASER-03}) and
there is one of these for each of the screen detecting sites $A_{2}^{i}$.

Next, we construct the generalized POVM elements $E_{2,0}^{i}$ from the
generalized Kraus operators. By definition, $E_{2,0}^{i}\equiv \bar{M}%
_{2,0}^{i}M_{2,0}^{i}$, where there is no sum over $i$. We find%
\begin{equation}
E_{2,0}^{i}=s_{0}^{1}\bar{s}_{0}^{1}\sum_{a,b=1}^{2}\bar{\alpha}^{a}\alpha
^{b}\bar{V}^{i,a}V^{i,b}A_{0}^{1}\bar{A}_{0}^{1}.
\end{equation}%
By construction these are all positive operators. As a check on the physical
correctness of our formalism, we can use the semi-unitary relations (\ref%
{ERASER-02}) to show that
\begin{equation}
\sum_{i=1}^{S}E_{2,0}^{i}=s_{0}^{1}\bar{s}_{0}^{1}A_{0}^{1}\bar{A}%
_{0}^{1}\equiv I_{0}^{EFF},
\end{equation}%
where $I_{0}^{EFF}$ is the effective identity operator for stage $\Omega
_{0} $.

The outcome signal rates $\Pr (A_{2}^{i}|\Psi _{0})$ are defined by $\Pr
(A_{2}^{i}|\Psi _{0})\equiv \bar{\Psi}_{0}E_{2,0}^{i}\Psi _{0}$ and are
found to be%
\begin{equation}
\Pr (A_{2}^{i}|\Psi _{0})=|\Psi _{0}^{1}|^{2}\sum_{a,b=1}^{2}\bar{\alpha}%
^{a}\alpha ^{b}\bar{V}^{i,a}V^{i,b}.
\end{equation}%
These rates contain the quantum interference contributions expected from
quantum principles. Using the semi-unitarity relations (\ref{ERASER-02}),
these rates add up as expected to the beam production rate $|\Psi
_{0}^{1}|^{2}$, as required from probability conservation.

This completes our analysis of the double slit experiment.

\begin{center}
\begin{figure}[t!]
\centerline{\psfig{file=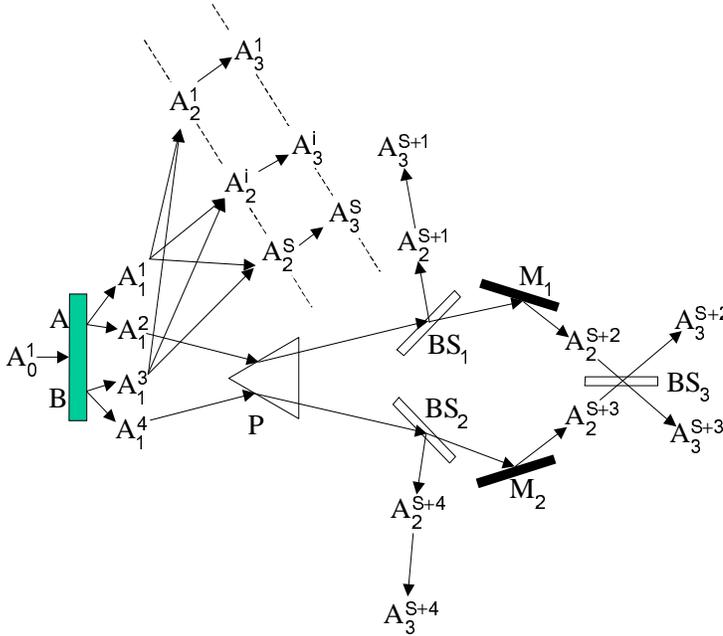,width=4in}}
\caption{The delayed-choice quantum eraser experiment}
\end{figure}
\end{center}

\section{Delayed-choice quantum eraser}

We turn now to the delayed-choice quantum eraser. The basic experiment is
shown in Figure $2$ \cite{KIM+AL-2000}. A photon source $A_{0}^{1}$ sends a
beam onto a crystal which is arranged to produce a superposition of coherent
photon pairs associated with sites $A$ and $B$ as shown. One component of
each pair is collimated and passed onto a screen containing detectors $%
A_{2}^{1},A_{2}^{2},\ldots ,A_{2}^{S}$, exactly as for the DS experiment
discussed in the previous section. In effect, positions $A$ and $B$ act as a
pair of slits for a DS experiment.

In this case, however, the situation is more complicated. For each pair, one
component is passed into the DS screen whilst the other passes into a prism $%
P$ which deflects it onto a beam-splitter. Component $A_{1}^{2}$ is passed
onto beam-splitter $BS_{1}$ whilst component $A_{1}^{4}$ is passed onto
beam-splitter $BS_{2}$. Each of these beam-splitters acts on its own
incident beam and splits it further into two. In each case, the reflected
component is passed onto a photon detector, either at $A_{3}^{S+1}$ or at $%
A_{3}^{S+4}$, whilst the transmitted component is passed onto a third
beam-splitter $BS_{3}$, where interference takes place, with subsequent
detection at $A_{3}^{S+2}$ and $A_{3}^{S+3}.$

Various discussions of this arrangement suggest that choices made by the
experimentalist at $BS_{3}$ can influence the interference patterns seen on
the screen containing $A_{2}^{1},\ldots ,A_{2}^{S}$, even though the signals
in that screen may have been captured much earlier.

We proceed with our stage analysis as follows. As with the DS experiment, we
represent our initial source at stage $\Omega _{0}$ by the labstate $\Psi
_{0}=\Psi _{0}^{1}s_{0}^{1}A_{0}^{1}$.

The next stage $\Omega _{1\text{ }}$is defined by the production of two
correlated photon pairs. These pairs are each assumed spinless. The
formalism can readily deal with any situation where this is not the case.
The first pair is generated at point $A$ on the crystal whilst the other
pair is generated at point $B$. The transformation from $\Omega
_{0}\rightarrow \Omega _{1}$ is given by%
\begin{eqnarray}
U_{1,0}s_{0}^{1}A_{0}^{1} &=&\alpha \frac{1}{\sqrt{2}}%
[s_{1}^{1,L}s_{1}^{2,R}+s_{1}^{1,R}s_{1}^{2,L}]A_{1}^{1}A_{1}^{2}+  \notag \\
&&\beta \frac{1}{\sqrt{2}}%
[s_{1}^{1,L}s_{1}^{2,R}+s_{1}^{1,R}s_{1}^{2,L}]A_{1}^{3}A_{1}^{4}
\end{eqnarray}%
where $|\alpha |^{2}+|\beta |^{2}=1$ and $s_{1}^{1,L}$ represents the spin
state of a left-handed circularly polarized photon moving along direction $%
A_{1}^{1}$, and so on. Here terms such as $A_{n}^{i}A_{n}^{j}$ represent the
two-signal labstate $\mathbb{A}_{i,n}^{+}\mathbb{A}_{j,n}^{+}|0,n)$ \cite%
{J2007C}. Hence the effective transition operator is%
\begin{equation}
U_{1,0}\backsimeq s_{1}^{1}\bar{s}_{0}^{1}[\alpha A_{1}^{1}A_{1}^{2}+\beta
A_{1}^{3}A_{1}^{4}]\bar{A}_{0}^{1},
\end{equation}%
where%
\begin{equation}
s_{n}^{1}\equiv \frac{1}{\sqrt{2}}%
[s_{n}^{1,L}s_{n}^{2,R}+s_{n}^{1,R}s_{n}^{2,L}]
\end{equation}%
represents an entangled two photon state of total angular momentum zero at
stage $\Omega _{n}$, $n=1,2,3$. Although individual photon wave components
get channeled into four possible directions as shown, the internal total
angular moment state remains unaffected during this particular experiment.

The next stage change is from $\Omega _{1}$ to $\Omega _{2}$ and given by%
\begin{align}
U_{2,1}s_{1}^{1}A_{1}^{1}A_{1}^{2}&
=s_{2}^{1}\sum_{i=1}^{S}V^{i,A}A_{2}^{i}\{t_{1}A_{2}^{S+2}+ir_{1}A_{2}^{S+1}%
\}  \notag \\
U_{2,1}s_{1}^{1}A_{1}^{3}A_{1}^{4}&
=s_{2}^{1}\sum_{i=1}^{S}V^{i,B}A_{2}^{i}\{t_{2}A_{2}^{S+3}+ir_{2}A_{2}^{S+4}%
\}.
\end{align}%
Here, one component beam from each pair is focused on the detecting screen
whilst the other component is channeled onto either beam-splitter $BS_{1}$
or $BS_{2}$, as shown. The $\{V^{i,A}\}$ represents the amplitudes for
landing on the screen from the pair sourced from point $A$, and similarly
for the pair sourced from point $B$. The coefficients $t_{i},r_{i}$ are
characteristic transmission and reflection parameters associated with $%
BS_{i} $. It is very useful not to set these parameters to the conventional
value $1/\sqrt{2}$ but to keep them open and available to be changed. It is
in these parameters that we shall encode the observer's freedom of choice in
this particular experiment.

From the above we find the effective transition operator
\begin{align}
U_{2,1}& \backsimeq s_{2}^{1}\bar{s}_{1}^{1}\sum_{i=1}^{S}A_{2}^{i}[V^{i,A}%
\{t_{1}A_{2}^{S+2}+ir_{1}A_{2}^{S+1}\}\bar{A}_{1}^{1}\bar{A}_{1}^{2}  \notag
\\
& \ \ \ \ \ \ +V^{i,B}\{t_{2}A_{2}^{S+3}+ir_{2}A_{2}^{S+4}\}\bar{A}_{1}^{3}%
\bar{A}_{1}^{4}].
\end{align}%
The final transition is from stage $\Omega _{2}$ to $\Omega _{3}$ and
involves four terms:%
\begin{align}
U_{3,2}s_{2}^{1}A_{2}^{i}A_{2}^{S+1}& =s_{3}^{1}A_{3}^{i}A_{3}^{S+1},  \notag
\\
U_{3,2}s_{2}^{1}A_{2}^{i}A_{2}^{S+2}&
=s_{3}^{1}A_{3}^{i}\{t_{3}A_{3}^{S+3}+ir_{3}A_{3}^{S+2}\}  \notag \\
U_{3,2}s_{2}^{1}A_{2}^{i}A_{2}^{S+3}&
=s_{3}^{1}A_{3}^{i}\{t_{3}A_{3}^{S+2}+ir_{3}A_{3}^{S+3}\} \\
U_{3,2}s_{2}^{1}A_{2}^{i}A_{2}^{S+4}& =s_{3}^{1}A_{3}^{i}A_{3}^{S+4}.  \notag
\end{align}%
This gives
\begin{align}
U_{3,2}& \backsimeq s_{3}^{1}\bar{s}_{2}^{1}\sum_{i=1}^{S}A_{3}^{i}\bar{A}%
_{2}^{i}[A_{3}^{S+1}\bar{A}_{2}^{S+1}+\{t_{3}A_{3}^{S+3}+ir_{3}A_{3}^{S+2}\}%
\bar{A}_{2}^{S+2}  \notag \\
& \ \ \ \ \ \ +\{t_{3}A_{3}^{S+2}+ir_{3}A_{3}^{S+3}\}\bar{A}%
_{2}^{S+3}+A_{3}^{S+4}\bar{A}_{2}^{S+4}].
\end{align}%
The complete evolution operators is given by $U_{3,0}\equiv
U_{3,2}U_{2,1}U_{1,0}$. Using the above results we find%
\begin{equation}
U_{3,0}\backsimeq s_{3}^{1}\bar{s}_{0}^{1}\sum_{i=1}^{S}A_{3}^{i}\left\{
\begin{array}{c}
\alpha V^{i,A}ir_{1}A_{3}^{S+1}+\beta V^{i,B}ir_{2}A_{3}^{S+4} \\
+[ir_{3}V^{i,A}t_{1}\alpha +t_{3}V^{i,B}t_{2}\beta ]A_{3}^{S+2} \\
+[t_{3}V^{i,A}t_{1}\alpha +ir_{3}V^{i,B}t_{2}\beta ]A_{3}^{S+3}\}%
\end{array}%
\right\} \bar{A}_{0}^{1}\text{.}
\end{equation}

There are four Kraus operators associated with each detector on the screen,
each of the form%
\begin{equation}
M_{3,0}^{i,S+k}\equiv \bar{A}_{3}^{i}\bar{A}_{3}^{S+k}U_{3,0},\ \ \
i=1,2,\ldots ,S,\ \ \ k=1,2,3,4,
\end{equation}%
We find%
\begin{align}
M_{3,0}^{i,S+1}& =s_{3}^{1}\bar{s}_{0}^{1}\alpha V^{i,A}ir_{1}\bar{A}%
_{0}^{1},  \notag \\
M_{3,0}^{i,S+2}& =s_{3}^{1}\bar{s}_{0}^{1}[ir_{3}V^{i,A}t_{1}\alpha
+t_{3}V^{i,B}t_{2}\beta ]\bar{A}_{0}^{1},  \notag \\
M_{3,0}^{i,S+3}& =s_{3}^{1}\bar{s}_{0}^{1}[t_{3}V^{i,A}t_{1}\alpha
+ir_{3}V^{i,B}t_{2}\beta ]\bar{A}_{0}^{1}, \\
M_{3,0}^{i,S+4}& =s_{3}^{1}\bar{s}_{0}^{1}\beta V^{i,B}ir_{2}\bar{A}_{0}^{1}.
\notag
\end{align}%
These give four POVMs associated with each detector on the screen:
\begin{eqnarray}
E_{3,0}^{i,S+1} &\equiv &\bar{M}_{3,0}^{i,S+1}M_{3,0}^{i,S+1}=r_{1}^{2}|%
\alpha |^{2}|V^{i,A}|^{2}s_{0}^{1}\bar{s}_{0}^{1}A_{0}^{1}\bar{A}_{0}^{1},
\notag \\
E_{3,0}^{i,S+2} &\equiv &\bar{M}%
_{3,0}^{i,S+2}M_{3,0}^{i,S+2}=|ir_{3}V^{i,A}t_{1}\alpha
+t_{3}V^{i,B}t_{2}\beta |^{2}s_{0}^{1}\bar{s}_{0}^{1}A_{0}^{1}\bar{A}%
_{0}^{1},  \notag \\
E_{3,0}^{i,S+3} &\equiv &\bar{M}%
_{3,0}^{i,S+3}M_{3,0}^{i,S+3}=|t_{3}V^{i,A}t_{1}\alpha
+ir_{3}V^{i,B}t_{2}\beta |^{2}s_{0}^{1}\bar{s}_{0}^{1}A_{0}^{1}\bar{A}%
_{0}^{1},  \notag \\
E_{3,0}^{i,S+4} &=&\bar{M}_{3,0}^{i,S+4}M_{3,0}^{i,S+4}=r_{2}^{2}|\beta
|^{2}|V^{i,B}|^{2}s_{0}^{1}\bar{s}_{0}^{1}A_{0}^{1}\bar{A}_{0}^{1}.  \notag
\\
&&
\end{eqnarray}%
It is straightforward to check that
\begin{equation}
\sum_{i=1}^{S}\sum_{k=1}^{4}E_{3,0}^{i,S+k}=s_{0}^{1}\bar{s}_{0}^{1}A_{0}^{1}%
\bar{A}_{0}^{1}=I_{0}^{Eff},
\end{equation}%
the effective identity operator for the initial stage Hilbert space.

There are four coincidence rates $\Pr (A_{3}^{i}A_{3}^{S+k}|\Psi _{0})$
associated with each detector on the screen, involving one of the detectors $%
A_{3}^{S+k}$, $k=1,2,3,4$. These rates are defined by%
\begin{equation}
\Pr (A_{3}^{i}A_{3}^{S+k}|\Psi _{0})\equiv \bar{\Psi}_{0}E_{3,0}^{i,S+k}\Psi
_{0}.
\end{equation}%
We find%
\begin{eqnarray}
\Pr (A_{3}^{i}A_{3}^{S+1}|\Psi _{0}) &=&|\Psi _{0}^{1}|^{2}r_{1}^{2}|\alpha
|^{2}|V^{i,A}|^{2},  \notag \\
\Pr (A_{3}^{i}A_{3}^{S+2}|\Psi _{0}) &=&|\Psi
_{0}^{1}|^{2}|ir_{3}V^{i,A}t_{1}\alpha +t_{3}V^{i,B}t_{2}\beta |^{2},  \notag
\\
\Pr (A_{3}^{i}A_{3}^{S+3}|\Psi _{0}) &=&|\Psi
_{0}^{1}|^{2}|t_{3}V^{i,A}t_{1}\alpha +ir_{3}V^{i,B}t_{2}\beta |^{2},  \notag
\\
\Pr (A_{3}^{i}A_{3}^{S+4}|\Psi _{0}) &=&|\Psi _{0}^{1}|^{2}r_{2}^{2}|\beta
|^{2}|V^{i,B}|^{2}.  \label{ERASER-06}
\end{eqnarray}

There are several observations to be made about these results.

\begin{enumerate}
\item The parameters $t_{i},r_{i}$ for beam-splitter $BS_{i}$ represent
places in the apparatus where the experimentalist could make changes, either
before or after signals have been registered on the screen during any given
run of the experiment. In other words, choices can be made at $BS_{1}$, $%
BS_{2}$ and $BS_{3}$ which affect various incidence rates. The question is,
does any change made by the experimentalist at any beam-splitter affect
anything that has been measured \emph{before} that change was made? In
particular, can any change in $BS_{3}$ affect what has already happened on
the screen?

By inspection of (\ref{ERASER-06}), we see that no change in $t_{3}$ or $%
r_{3}$, subject to $t_{3}^{2}+r_{3}^{2}=1,$ has any effect whatsoever on $%
\Pr (A_{3}^{i}A_{3}^{S+1}|\Psi _{0})$ or $\Pr (A_{3}^{i}A_{3}^{S+4}|\Psi
_{0})$. These coincidence rates actually involve signal detection completed
during \emph{earlier} stages. The conclusion therefore is that any
suggestion that delayed-choice can erase information acquired in the past is
false and misleading.

\item It is true that changes in $t_{3}$ and $r_{3}$ affect $\Pr
(A_{3}^{i}A_{3}^{S+2}|\Psi _{0})$ and $\Pr (A_{3}^{i}A_{3}^{S+3}|\Psi _{0})$%
. However no acausality is involved, because a coincidence rate is undefined
until signals from both detectors involved have been counted. $\Pr
(A_{3}^{i}A_{3}^{S+2}|\Psi _{0})$ and $\Pr (A_{3}^{i}A_{3}^{S+3}|\Psi _{0})$
cannot be measured until \emph{after} the choice of $t_{3}$ and $r_{3}$.

Suggestions that events in stage $\Omega _{3}$ could influence events in
earlier stages do not take into account the crucial role of \emph{%
post-selection} in such experiments. The proper way to understand what is
happening is to view the role of the four detectors $A_{3}^{S+i}$ as a
post-selection processing of data already accumulated on the screen.

\item If we look at the total counting rates at each of the four detectors $%
A_{3}^{S+k}\equiv \sum\limits_{i=1}^{S}\Pr (A_{3}^{i}A_{3}^{S+1}|\Psi _{0})$%
, $k=1,2,3,4$, we find%
\begin{eqnarray}
\Pr (A_{3}^{S+1}|\Psi _{0}) &=&|\Psi _{0}^{1}|^{2}r_{1}^{2}|\alpha |^{2},
\notag \\
\Pr (A_{3}^{S+2}|\Psi _{0}) &=&|\Psi
_{0}^{1}|^{2}\{t_{1}^{2}r_{3}^{2}|\alpha |^{2}+t_{2}^{2}t_{3}^{2}|\beta
|^{2}\},  \notag \\
\Pr (A_{3}^{S+3}|\Psi _{0}) &=&|\Psi
_{0}^{1}|^{2}\{t_{1}^{2}t_{3}^{2}|\alpha |^{2}+t_{2}^{2}r_{3}^{2}|\beta
|^{2}\},  \notag \\
\Pr (A_{3}^{S+4}|\Psi _{0}) &=&|\Psi _{0}^{1}|^{2}r_{2}^{2}|\beta |^{2},
\end{eqnarray}%
using the semi-unitarity of the $\{V^{i,A}\}$ and $\{V^{i,B}\}$
coefficients. Again, changes made at $BS_{3}$ would have no effect on $\Pr
(A_{3}^{S+1}|\Psi _{0})$ or $\Pr (A_{3}^{S+4}|\Psi _{0})$.

\item If we look at the total count rate for a given detector on the screen,
we find%
\begin{eqnarray}
\Pr (A_{3}^{i}|\Psi _{0}) &\equiv &\sum_{k=1}^{4}\Pr
(A_{3}^{i}A_{3}^{S+k}|\Psi _{0})  \notag \\
&=&|\Psi _{0}^{1}|^{2}\{|\alpha |^{2}|V^{i,A}|^{2}+|\beta
|^{2}|V^{i,B}|^{2}\},
\end{eqnarray}%
which shows that no changes at any of the beam-splitters affects the pattern
observed on the screen.

\item Significantly, changes made at either $BS_{1}$ and/or $BS_{2}$ would
have an effect on the counting rates at $A_{3}^{S+2}$ and$A_{3}^{S+3}$. That
is physically possible because $BS_{1}$ and $BS_{2}$ are involved in stage $%
\Omega _{2}$, which is earlier than $\Omega _{3}$.

\item This particular experiment is a good one to illustrate the concept of
\emph{stage}. None of the detectors $A_{n}^{i}$ is assumed to have an
enduring identity throughout time. In that sense, they do not represent the
devices \emph{per se} constructed in a laboratory, which usually persist as
physical objects during many separate runs of an experiment. Rather, the $%
A_{n}^{i}$ represent a potential for information transfer between the
observer and the apparatus in stage $\Omega _{n}$. As discussed in our
account of the Franson-Bell experiment \cite{J2008E}, \emph{context} in the
form of which-path information can determine the dynamics of the $A_{n}^{i}$.

The observer, who is controlling the apparatus, has the freedom to decide
whether or not to actually look at a given detector at any given time to see
if a photon has been registered or not. In the quantum eraser experiment, $%
A_{3}^{1},A_{3}^{2},A_{3}^{3}$ and $A_{3}^{4}$ are all in the same stage,
even though their individual actual laboratory times could be very different.

What is most remarkable about quantum processes is that \emph{if} a signal
is not observed at a given detector at a given time, that detector can act
as a source for signals observed later on. Moreover, quantum rules tells us
to add signal amplitudes whenever several such detectors are involved, as in
the DS experiment and then take the square modulus in order to calculate
relative probabilities.
\end{enumerate}

\section{Which-path measure}

The double-slit and eraser experiments discussed above belong to an
important class of experiment which, to use colloquial terminology, provide
partial or complete information about which path a photon had taken in its
journey from initial to final stages. Another important experiment which
belongs to this category is the Franson-Bell experiment, which we have
recently discussed \cite{J2008E}. We shall discuss below another
example,Wheeler's delayed-choice experiment.

Each of these experiments carries with it contextual attributes arising from
the experimental setup which determine the extent to which paths can be
determined from the data or not. For example, the double-slit experiment
with both slits open gives zero information about which slit a particular
detected photon came from. On the other hand, the same setup with one of the
slits blocked up gives us total information as to where any of the detected
photons originated.

It is of interest therefore to find some measure or parameter $\Phi $ which
is characteristic of any given experimental setup and which gives us an
indication as to how much which-path information we could extract. In the
absence of any deeper analysis, possibly based on entropic grounds, our
choice is to define $\Phi $ as the total probability of determining for sure
full path information from a single detected photon, i.e.,
\begin{equation}
\Phi \equiv Prob(\text{full path information}|\text{single photon anywhere}).
\end{equation}%
In the case of the double-slit experiment discussed above we find $\Phi
_{DS}=0$. On the other hand, in the case of the delayed-choice quantum
eraser discussed above, we find%
\begin{equation}
\Phi =\frac{\Pr (A_{3}^{S+1}|\Psi _{0})+\Pr (A_{3}^{S+4}|\Psi _{0})}{|\Psi
_{0}^{1}|^{2}}=r_{1}^{2}|\alpha |^{2}+r_{2}^{2}|\beta |^{2}.
\end{equation}%
In the conventional symmetric situation when $r_{1}=r_{2}=1/\sqrt{2}$, $\Phi
=1/2$ as we should expect. When $r_{1}=r_{2}=0$, any single photon detected
in an off-screen detector would occur only either in $A_{3}^{S+2}$ or else $%
A_{3}^{S+3}$ and no path information could normally be obtained. However,
there is a pathology in this case, because $r_{1}=r_{2}=0$ gives $\Pr
(A_{3}^{S+2}|\Psi _{0})=|\Psi _{0}^{1}|^{2}\{r_{3}^{2}|\alpha
|^{2}+t_{3}^{2}|\beta |^{2}\}$ and $\Pr (A_{3}^{S+3}|\Psi _{0})=|\Psi
_{0}^{1}|^{2}\{t_{3}^{2}|\alpha |^{2}+r_{3}^{2}|\beta |^{2}\}$. If the
experimentalist had set $t_{3}=0$ or $r_{3}=0$, then a single photon
detected at $A_{3}^{S+2}$ or $A_{3}^{S+3}$ would now give information about
which path had been taken. Of course, this is equivalent to having no
beam-splitters and is therefore of limited value.

In the next section we shall discuss Wheeler's delayed-choice experiment and
determine the which-path parameter for it.

\begin{figure}[t!]
\centerline{\psfig{file=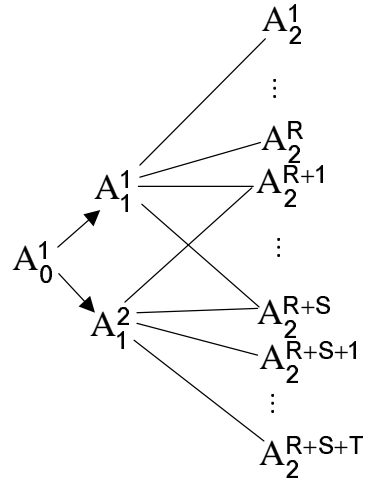,width=2in}}
\caption{Idealization of Wheeler's delayed-choice experiment}
\end{figure}

\section{Wheeler's delayed-choice experiment}

Wheeler's delayed-choice experiment can be regarded as a double-slit
experiment with a modified screen. Some of the detectors can receive quantum
signals from both slits, whilst the others can receive a signal from only
one of the slits. The interest this experiment generates comes from the
possibility that the observer can decide in principle which detector
receives which signal/s \emph{after} light has left the two slits. A recent
experiment which confirms quantum expectation was done with a Mach-Zehnder
interferometer, such that the final beam-splitter could be removed whilst
the light was on its way from the first beam-splitter \cite{JACQUES+AL-2006}.

An idealized version of this experiment is shown in Figure $3$. The details
are much the same as the DS experiment studied first, but with the
difference that now there are three groups of detectors on the screen. $%
A_{2}^{1}$ to $A_{2}^{R}$ can each receive a quantum amplitude from $%
A_{1}^{1}$ only, $A_{2}^{R+1}$ to $A_{2}^{R+S}$ can each receive quantum
amplitudes from $A_{1}^{1}$ and from $A_{1}^{2}$, and $A_{2}^{R+S+1}$ to $%
A_{2}^{R+S+T}$ can each receive a quantum amplitude from $A_{1}^{2}$ only.
We can imagine that the experimentalist can shuffle the values of $R$, $S$
and $T$ during any given run \emph{after} they were sure that light had left
the two slits and \emph{before} any impact on the screen. Of course, any
actual experiment would require a lot of analysis of the data,
post-selecting signals corresponding to equivalent values of $R$, $S$ and $T$%
.

We can encode the dynamics by writing%
\begin{equation}
U_{2,1}s_{1}^{1}A_{1}^{a}=s_{2}^{1}\sum_{i=1}^{R+S+T}V^{i,a}A_{2}^{i},\ \ \
a=1,2,
\end{equation}%
with the condition that
\begin{equation}
V^{i,1}=0,\ \ \ \ \ i>R+S,\ \ \ \ \ V^{i,2}=0,\ \ \ \ \ i\leqslant R\text{.}
\end{equation}%
The semi-unitarity relations are then equivalent to%
\begin{equation}
\sum_{i=1}^{R+S}|V^{i,1}|^{2}=1,\ \ \ \sum_{i=R+1}^{R+S}\bar{V}%
^{i,1}V^{i,2}=0,\ \ \ \sum_{i=R+S+1}^{R+S+T}|V^{i,2}|^{2}=1.
\end{equation}

Applying the results found for the DS experiment we find%
\begin{eqnarray}
\Pr (A_{2}^{i}|\Psi _{0}) &=&|\Psi _{0}^{1}|^{2}|\alpha ^{1}V^{i,1}|^{2},\ \
\ 1\leqslant i\leqslant R,  \notag \\
P(A_{2}^{i}|\Psi _{0}) &=&|\Psi _{0}^{1}|^{2}\sum_{a,b=1}^{2}\bar{\alpha}%
^{a}\alpha ^{b}\bar{V}^{i,a}V^{i,b},\ \ \ R<i\leqslant R+S,  \notag \\
P(A_{2}^{i}|\Psi _{0}) &=&|\Psi _{0}^{1}|^{2}|\alpha ^{2}V^{i,2}|,\ \ \
R+S<i\leqslant R+S+T.  \notag \\
&&
\end{eqnarray}%
From this we find the which-path parameter to be%
\begin{equation}
\Phi =|\alpha ^{1}|^{2}\sum_{i=1}^{R}|V^{i,1}|^{2}+|\alpha
^{2}|^{2}\sum_{i=R+S+1}^{R+S+T}|V^{i,2}|.
\end{equation}%
This reduces to unity when $S=0$ as expected and zero when both $R$ and $T$
are zero.

The recent experiment of Jacques et al \cite{JACQUES+AL-2006} is equivalent
to the above scenario with $R+S+T=2$; the configuration with the second
beam-splitter removed corresponds to $R=T=1$, $S=0$, whilst that with the
second beam-splitter in operation corresponds to $R=T=0$, $S=2$.

\section{The double-slit quantum eraser}

The above experiments have not involved photon spin significantly. The
experiment we discuss next requires a careful analysis of spin.

Prior to the delayed-choice quantum eraser experiment of Jacques et al \cite%
{JACQUES+AL-2006}, the double-slit quantum eraser experiment of Walborn et
al \cite{WALBORN-2002} had demonstrated the empirical validity of the stage
concept in quantum mechanics. Their experiment is discussed in two parts.

\subsection{No polarization control}

The first part of the experiment is shown in Figure $4$. A spinless photon
pair is produced, with one photon $s$ passed onto a double-slit and then
onto a screen, whilst the other photon $p$ is passed onto a detector.
Coincidence measurements are taken involving fixed position screen impacts
and $p$ photon detection, with no polarization input involved.

\begin{figure}[t!]
\centerline{\psfig{file=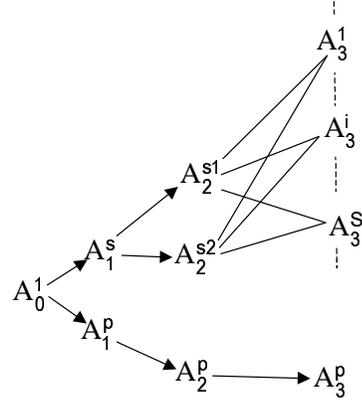,width=2in}}
\caption{Double-slit quantum eraser without polarization control.}
\end{figure}

With an initial state $\Psi _{0}=\Psi _{0}^{1}s_{0}^{1}A_{0}^{1}$, the
evolution from $\Omega _{0}\rightarrow \Omega _{1}$ is given by the operator%
\begin{equation}
U_{1,0}=\dfrac{1}{\sqrt{2}}\{s_{1}^{sH}s_{1}^{pV}+s_{1}^{sV}s_{1}^{pH}\}\bar{%
s}_{0}^{1}A_{1}^{s}A_{1}^{p}\bar{A}_{0}^{1},
\end{equation}%
where $H$ and $V$ are the horizontal and vertical polarization degrees of
freedom. The next step is $\Omega _{1}\rightarrow \Omega _{2}$, with
evolution operator%
\begin{equation}
U_{2,1}=\sum\limits_{a=1}^{2}\alpha ^{a}\{s_{2}^{sH}s_{2}^{pV}\bar{s}%
_{1}^{sH}\bar{s}_{1}^{pV}+s_{2}^{sV}s_{2}^{pH}\bar{s}_{1}^{sV}\bar{s}%
_{1}^{pH}\}A_{2}^{sa}A_{2}^{p}\bar{A}_{1}^{s}\bar{A}_{1}^{p},
\end{equation}%
where as with the basic double slit experiment discussed in $\S II$, $%
|\alpha ^{1}|^{2}+|\alpha ^{2}|^{2}=1$.

The final stage transition $\Omega _{2}\rightarrow \Omega _{3}$ is described
by evolution operator%
\begin{equation}
U_{3,2}=\sum\limits_{a=1}^{2}\sum\limits_{i=1}^{S}V^{i,a}\left\{
\begin{array}{c}
s_{3}^{sH}s_{3}^{pV}\bar{s}_{2}^{sH}\bar{s}_{2}^{pV}+ \\
s_{3}^{sV}s_{3}^{pH}\bar{s}_{2}^{sV}\bar{s}_{2}^{pH}%
\end{array}%
\right\} A_{3}^{i}A_{3}^{p}\bar{A}_{2}^{sa}\bar{A}_{2}^{p},
\end{equation}%
where the screen is assumed to consist of $S$ detector sites and the $%
\{V^{i,a}\}$ coefficients satisfy the semi-unitarity conditions (\ref%
{ERASER-02}). The complete evolution operator is given by
\begin{equation}
U_{3,0}=\dfrac{1}{\sqrt{2}}\sum\limits_{a=1}^{2}\sum\limits_{i=1}^{S}\alpha
^{a}V^{i,a}\{s_{3}^{sH}s_{3}^{pV}+s_{3}^{sV}s_{3}^{pH}\}\bar{s}%
_{0}^{1}A_{3}^{i}A_{3}^{p}\bar{A}_{0}^{1},
\end{equation}%
which gives the Kraus operators%
\begin{equation}
M_{3,0}^{i,p}=\dfrac{1}{\sqrt{2}}\sum\limits_{a=1}^{2}\alpha
^{a}V^{i,a}\{s_{3}^{sH}s_{3}^{pV}+s_{3}^{sV}s_{3}^{pH}\}\bar{s}_{0}^{1}\bar{A%
}_{0}^{1}.
\end{equation}%
These give the POVMs%
\begin{equation}
E_{3,0}^{i,p}=\sum\limits_{a,b=1}^{2}\bar{\alpha}^{b}\alpha ^{a}\bar{V}%
^{i,b}V^{i,a}s_{0}^{1}A_{0}^{1}\bar{s}_{0}^{1}\bar{A}_{0}^{1},\ \ \
i=1,2,\ldots ,S
\end{equation}%
from which we find the coincidence rates%
\begin{equation}
\Pr (A_{3}^{i}A_{3}^{p}|\Psi _{0})=|\Psi _{0}^{1}|^{2}\sum\limits_{a,b=1}^{2}%
\bar{\alpha}^{b}\alpha ^{a}\bar{V}^{i,b}V^{i,a}.
\end{equation}%
These demonstrate double-slit interference, because detection of the $p$
photon provides no which-way information.

\subsection{Polarization control}

\begin{figure}[t]
\centerline{\psfig{file=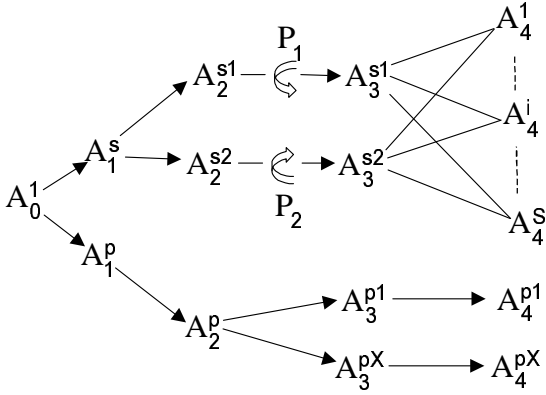,width=3in}}
\caption{Double-slit quantum eraser with polarization control.}
\end{figure}

The experiment is now repeated with some modifications, shown schematically
in Figure $5$. Two quarter-wavelength polarizers $P_{1}$ and $P_{2}$ are
introduced, $P_{1}$ in front of slit $1$ and $P_{2}$ in front of slit $2$.
These polarizers have equal and opposite effects, given by

\begin{align}
s^{1H}& \rightarrow s^{1L},\ \ \ s^{1V}\rightarrow is^{1R},  \notag \\
s^{2H}& \rightarrow s^{2R},\ \ \ s^{2V}\rightarrow -is^{2L},
\end{align}%
where $H$, $V$ represent horizontal and vertical plane polarizations, whilst
$R$, $L$ represent right-handed and left-handed circularly polarized states.
In addition, the observer can insert a plane polarizer in front of the $p$
photon detector. Our formalism deals with this as if this were a choice.

The details of the evolution are the same as for the unpolarized situation
up to stage $\Omega _{2}$, i.e., we have%
\begin{equation}
U_{2,0}=\dfrac{1}{\sqrt{2}}\{s_{2}^{sH}s_{2}^{pV}+s_{2}^{sV}s_{2}^{pH}\}\sum%
\limits_{a=1}^{2}\alpha ^{a}A_{2}^{sa}A_{2}^{p}\bar{s}_{0}^{1}\bar{A}%
_{0}^{1}.
\end{equation}%
In the next step from $\Omega _{2}\rightarrow \Omega _{3}$ we first
transform to circularly polarized states and then rewrite them in terms of
the linear polarization vectors $|+\rangle $ and $|-\rangle $, defined by%
\begin{equation}
|R\rangle =\frac{(1-i)}{2}\{|+\rangle +i|-\rangle \},\ \ \ |L\rangle =\frac{%
(1-i)}{2}\{i|+\rangle +|-\rangle \}.
\end{equation}%
Our conventions are exactly those in \cite{WALBORN-2002}. Then we find%
\begin{eqnarray}
U_{3,2}s_{2}^{sH}s_{2}^{pV}A_{2}^{s1}A_{2}^{p}
&=&s_{3}^{sL}s_{3}^{pV}A_{3}^{s1}A_{3}^{p}  \notag \\
&=&\frac{(1-i)}{2\sqrt{2}}\{is_{3}^{s+}+s_{3}^{s-}\}A_{3}^{s1}\times  \notag
\\
&&\{s_{3}^{p+}A_{3}^{p1}-s_{3}^{p-}A_{3}^{pX}\},
\end{eqnarray}%
etc., where the label $X$ indicates a choice. If $X=1$, that corresponds to
no polarizer placed in front of the detector of the $p$ photon, whereas $X=2$
corresponds to an ability to detect two possible polarizations at that
detector. The result is%
\begin{align}
U_{3,2}& =\frac{(1-i)}{2\sqrt{2}}\left(
\begin{array}{c}
\left\{ s_{3}^{p+}A_{3}^{p1}-s_{3}^{p-}A_{3}^{pX}\right\} \times \ \ \ \ \ \
\ \ \ \ \ \ \ \ \ \ \ \ \ \ \ \ \ \  \\
\left[
\begin{array}{c}
\{is_{3}^{s+}+s_{3}^{s-}\}A_{3}^{s1}\bar{A}_{2}^{s1}+ \\
\{s_{3}^{s+}+is_{3}^{s-}\}A_{3}^{s2}\bar{A}_{2}^{s2}%
\end{array}%
\right] \bar{s}_{2}^{sH}\bar{s}_{2}^{pV}\bar{A}_{2}^{p}+%
\end{array}%
\right.  \notag \\
& \ \ \ \ \ \ \ \ \ \ \ \ \ \ \ \ \ \ \ \left.
\begin{array}{c}
\left\{ s_{3}^{p+}A_{3}^{p1}+s_{3}^{p-}A_{3}^{pX}\right\} \times \ \ \ \ \ \
\ \ \ \ \ \ \ \ \ \ \ \ \ \ \ \ \  \\
\left[
\begin{array}{c}
\{is_{3}^{s+}-s_{3}^{s-}\}A_{3}^{s1}\bar{A}_{2}^{s1}+ \\
\{s_{3}^{s+}-is_{3}^{s-}\}A_{3}^{s2}\bar{A}_{2}^{s2})%
\end{array}%
\right] \bar{s}_{2}^{sV}\bar{s}_{2}^{pH}\bar{A}_{2}^{p}%
\end{array}%
\right) .
\end{align}

The final evolution operator $U_{4,3}$ involves screen impacts and is given
by%
\begin{eqnarray}
U_{4,3}
&=&\sum\limits_{a=1}^{2}\sum\limits_{i=1}^{S}V^{i,a}A_{4}^{i}\{s_{4}^{s+}%
\bar{s}_{3}^{s+}+s_{4}^{s-}\bar{s}_{3}^{s-}\}\times  \notag \\
&&\lbrack s_{4}^{p+}\bar{s}_{3}^{p+}A_{4}^{p1}\bar{A}_{3}^{p1}+s_{4}^{p-}%
\bar{s}_{3}^{p-}A_{4}^{pX}\bar{A}_{3}^{pX}]\bar{A}_{3}^{sa}.
\end{eqnarray}

Hence we find%
\begin{eqnarray}
U_{4,0} &=&\dfrac{(1-i)}{2}\sum\limits_{i=1}^{S}A_{4}^{i}\times
\label{ERASER-10} \\
&&\left\{
\begin{array}{c}
\alpha
^{1}V^{i,1}[is_{4}^{s+}s_{4}^{p+}A_{4}^{p1}-s_{4}^{s-}s_{4}^{p-}A_{4}^{pX}]+
\\
\alpha
^{2}V^{i,2}[s_{4}^{s+}s_{4}^{p+}A_{4}^{p1}-is_{4}^{s-}s_{4}^{p-}A_{4}^{pX}]%
\end{array}%
\right\} \bar{s}_{0}^{1}\bar{A}_{0}^{1}.  \notag
\end{eqnarray}

We are now in a position to make a choice as to what happens at the $p$
detector. First, we remove the $p$ detector polarizer.

\subsection{Case I: no erasure}

For this scenario, we take $X=1$. Then the total evolution operator is%
\begin{eqnarray}
U_{4,0} &=&\dfrac{(1-i)}{2}\sum\limits_{i=1}^{S}A_{4}^{i}A_{4}^{p}\times  \\
&&\ \ \left\{
\begin{array}{c}
\alpha ^{1}V^{i,1}[is_{4}^{s+}s_{4}^{p+}-s_{4}^{s-}s_{4}^{p-}]+ \\
\alpha ^{2}V^{i,2}[s_{4}^{s+}s_{4}^{p+}-is_{4}^{s-}s_{4}^{p-}]%
\end{array}%
\right\} \bar{s}_{0}^{1}\bar{A}_{0}^{1}.  \notag
\end{eqnarray}%
This gives the Kraus operators%
\begin{equation}
M_{4,0}^{i,p}=\dfrac{(1-i)}{2}\left\{
\begin{array}{c}
\alpha ^{1}V^{i,1}[is_{4}^{s+}s_{4}^{p+}-s_{4}^{s-}s_{4}^{p-}]+ \\
\alpha ^{2}V^{i,2}[s_{4}^{s+}s_{4}^{p+}-is_{4}^{s-}s_{4}^{p-}]%
\end{array}%
\right\} \bar{s}_{0}^{1}\bar{A}_{0}^{1},
\end{equation}%
from which we construct the POVMs%
\begin{equation}
E_{4,0}^{i,p}=\{|\alpha ^{1}V^{i,1}|^{2}+|\alpha
^{2}V^{i,2}|^{2}\}s_{0}^{1}A_{0}^{1}\bar{s}_{0}^{1}\bar{A}_{0}^{1}
\end{equation}%
Hence the coincidence rates involving screen site $i$ and the $p$ detector
are%
\begin{equation}
\Pr (A_{4}^{i}A_{4}^{p}|\Psi _{0})=|\Psi _{0}|^{2}\{|\alpha
^{1}V^{i,1}|^{2}+|\alpha ^{2}V^{i,2}|^{2}\},\ \ \ i=1,2,\ldots ,S,
\end{equation}%
which show no interference. This is precisely what was observed by Walborn
et al \cite{WALBORN-2002}. Essentially, placing $P_{1}$ and $P_{2}$ in front
of their respective slits destroys \emph{in principle }the lack of which-way
information so evident in the conventional unpolarized double-slit
experiment discussed ealier. In the current scenario, the experimentalist
could if so desired have determined the spin of every photon impacting on
the screen and thereby determine from which slit it had come. It is the mere
possibility of doing this that destroys the interference pattern.

\subsection{Case II: erasure: $X=2$}

Now we consider the effect of inserting a polarizing filter in front of the $%
p$ detector. In this case, the presence of a polarizing filter with variable
angle at the $p$ detector is equivalent to placing a Wollaston prism there
with two output beams with mutually orthogonal polarizations. In our
approach, this is described in terms of two detectors, $A_{4}^{p1}$ and $%
A_{4}^{p2}$, rather than one.

In this scenario, the total evolution operator is given by setting $X=2$ in\
(\ref{ERASER-10}):

\begin{eqnarray}
U_{4,0} &=&\dfrac{(1-i)}{2}\sum\limits_{i=1}^{S}A_{4}^{i}\times \\
&&\ \ \left\{
\begin{array}{c}
\alpha
^{1}V^{i,1}[is_{4}^{s+}s_{4}^{p+}A_{4}^{p1}-s_{4}^{s-}s_{4}^{p-}A_{4}^{p2}]+
\\
\alpha
^{2}V^{i,2}[s_{4}^{s+}s_{4}^{p+}A_{4}^{p1}-is_{4}^{s-}s_{4}^{p-}A_{4}^{p2}]%
\end{array}%
\right\} \bar{s}_{0}^{1}\bar{A}_{0}^{1}.  \notag
\end{eqnarray}

From this we find the Kraus operators%
\begin{align}
M_{4,0}^{i,p1}& =\dfrac{(1-i)}{2}s_{4}^{s+}s_{4}^{p+}\{i\alpha
^{1}V^{i,1}+\alpha ^{2}V^{i,2}\}\bar{s}_{0}^{1}\bar{A}_{0}^{1},  \notag \\
M_{4,0}^{i,p2}& =-\dfrac{(1-i)}{2}s_{4}^{s-}s_{4}^{p-}\{\alpha
^{1}V^{i,1}+i\alpha ^{2}V^{i,2}\}\bar{s}_{0}^{1}\bar{A}_{0}^{1}.
\end{align}%
and then the POVMs%
\begin{align}
E_{4,0}^{i,p1}& =\dfrac{1}{2}|i\alpha ^{1}V^{i,1}+\alpha
^{2}V^{i,2}|^{2}s_{0}^{1}A_{0}^{1}\bar{s}_{0}^{1}\bar{A}_{0}^{1}  \notag \\
E_{4,0}^{i,p2}& =\dfrac{1}{2}|\alpha ^{1}V^{i,1}+i\alpha
^{2}V^{i,2}|^{2}s_{0}^{1}A_{0}^{1}\bar{s}_{0}^{1}\bar{A}_{0}^{1}.
\end{align}%
Hence the two coincidence transition rate patterns are given by%
\begin{align}
\Pr (A_{3}^{i}A_{3}^{p1}|\Psi _{0})& =\dfrac{1}{2}|i\alpha
^{1}V^{i,1}+\alpha ^{2}V^{i,2}|^{2}|\Psi _{0}^{1}|^{2}  \notag \\
\Pr (A_{3}^{i}A_{3}^{p2}|\Psi _{0})& =\dfrac{1}{2}|\alpha
^{1}V^{i,1}+i\alpha ^{2}V^{i,2}|^{2}|\Psi _{0}^{1}|^{2},\ \ \ i=1,2,\ldots
,S.
\end{align}%
These now show interference, with one showing what would normally be
described as a \emph{fringe }pattern whilst the other showing an \emph{%
antifringe }pattern. Essentially, the insertion of the polarizer in front of
the $p$ photon detector erases the which-path information which previously
gave a non-interference pattern on the screen.

Most significantly, Walborn et al repeated the experiment with the screen
and $p$ photon detection order reversed with significant time differences
and found no change in the results. This is strong evidence for the validity
of the stages concept in such quantum process.

\section{Concluding remarks}

Our analysis supports the notion that quantum mechanics never actually
involves acausality. We should be worried if it did, for then our entire
view of what probability and information represent would need drastic
revision.

However, several factors would appear to conspire to make it look otherwise.
It is the case that some quantum interference experiments do suggest
acausality to the unwary. We have in mind here not only the delayed-choice
scenarios discussed here but also the Franson-Bell experiment \cite%
{FRANSON-1989,KWIAT-1993,J2008E}. In that experiment, the lack of which-path
information involves non-locality in time as well as non-locality in space
in a most spectacular fashion.

However, detailed analysis always reveals the basic fact that interference
phenomena arise from a lack of information about quantum \emph{states} and
have nothing specifically to do with the properties of particles per se. It
may be reasonable to talk about $``$photon self-interference$"$ when we know
we are dealing with one-signal experiments, but as the Franson-Bell
experiment and more recent ones demonstrate \cite{KIM-2003}, two-photon
state interference occurs under circumstances when individual photons simply
do not $``$overlap$"$.

Conceptual problems arise when our classical conditioning is relied on too
much. We would like to believe in photons as particles and we would like to
believe that time runs continuously. Both concepts have their uses, but
quantum mechanics requires a generalization of both. In the case of the
former, experiments tell us that we have to deal with interference of states
not particles. In the case of the latter, we cannot expect quantum processes
to evolve strictly according to an integrable timetable, such as coordinate
time, or even the physical time in a laboratory. What is important is
whether or not quantum information has been extracted. If it has been placed
$``$on hold$"$, as can be seen in our analysis of the delayed-choice eraser
and the double-slit eraser, then it can remain in a stage which could in
principle persist until the end of the universe. This is one way of
understanding unstable particles \cite{J2007G}: an undecayed unstable
particle is one trapped in an information bubble.

We end with a remarkable quote from experimentalists who have done real
experiments in this area \cite{JACQUES+AL-2006}:

$``$\emph{Once more, we find that Nature behaves in agreement with the
predictions of Quantum Mechanics even in surprising situations where a
tension with Relativity seems to appear.}$"$


\begin{thebibliography}{99}
\bibitem{WHEELER-1978} J. A. Wheeler, \emph{The }$``$\emph{past}$"$\emph{\
and the }$``$\emph{delayed-choice}$"$\emph{\ double-slit experiment} in
Mathematical Foundations of Quantum Theory, A.~R. Marlow, editor. Academic
Press (New York), 1978.

\bibitem{JACQUES+AL-2006} V. Jacques, E. Wu, F. Grosshans, F. Treussart, P.
Grangier, A. Aspect and J.-F. Roch, \emph{Science}, 315, 966, 2007.

\bibitem{WALBORN-2002} S. P. Walborn, M. O. TerraCunha, S. Padua and C. H.
Monken, \emph{Physical Review A}, 65, 033818, 1--6, 2002.

\bibitem{KIM+AL-2000} Y. H. Kim, R. Yu, S. P. Kulik, Y. H. Shih and M. O. Scully, \emph{Phys. Rev. Lett.}, 84, 1--5, 2000.

\bibitem{KIM-2003} Y. H. Kim, \emph{Physics Letters A}, 315, 352--355, 2003.

\bibitem{J2005A} J. Eakins and G. Jaroszkiewicz, \emph{A Quantum
Computational Approach to the Quantum Universe.} in New Developments in
Quantum Cosmology Research, A. Reimer, editor, volume 247 of \emph{Horizons
in World Physics}, chapter 1, pages 1--51, Nova Science Publishers, Inc. New
York, 2005.

\bibitem{J2007G} G. Jaroszkiewicz, \emph{J. Phys. A: Math. and Theor.}, 41,
095301, 2008.

\bibitem{J2007C} G. Jaroszkiewicz, \emph{Int. J. Mod. Phys. B}, 22(3), 123
-- 188, 2008.

\bibitem{FRANSON-1989} J. D. Franson, \emph{Phys. Rev. Lett.}, 62(19),
2205--2208, 1989.

\bibitem{KWIAT-1993} P. G. Kwiat, A. M. Steinberg and R. Y. Chiao, \emph{%
Physical Review A}, {47}(4), R2472--R2475, 1993.

\bibitem{J2008E} G. Jaroszkiewicz, \emph{Quantized detector network {POVMs}
and the Franson-Bell experiment}, arXiv:quant-ph-0809.3466, 1--9, 2008.
\end{thebibliography}
\end{document}